\documentclass[12pt]{article}
\usepackage{amsfonts}
\usepackage{bm}
\usepackage{graphicx}
\usepackage{multicol}

\begin{document}

\title{
Two-time correlation functions and the Lee-Yang zeros for an interacting Bose gas
}
\maketitle

\centerline {Kh. P. Gnatenko \footnote{E-Mail address: khrystyna.gnatenko@gmail.com}, A. Kargol \footnote{E-Mail address: akargol@hektor.umcs.lublin.pl}, V. M. Tkachuk \footnote{E-Mail address: voltkachuk@gmail.com}}
\medskip
\centerline {\small  $^{1,3}$ \it  Ivan Franko National University of Lviv, }
\centerline {\small \it Department for Theoretical Physics,12 Drahomanov St., Lviv, 79005, Ukraine}

 \centerline {  \small  $^{2}$ \it Instytut Matematyki, Uniwersytet Marii Curie-Sklodowskiej,}
\centerline {\small \it 20-031 Lublin, Poland}

\begin{abstract}
 Two-time correlation functions of a system of Bose particles are studied. We find relation of zeros of the correlation functions with the Lee-Yang zeros of partition function of the system.  Obtained relation gives the possibility to observe the  Lee-Yang zeros experimentally. A particular case of Bose particles on two levels is examined and zeros of two-time correlation functions and Lee-Yang zeros of partition function of the system are analyzed.

\end{abstract}

\section{Introduction}

In 1952 Lee and Yang  presented consideration that revolutionized the studies of phase transitions \cite{Yang52,Lee52}. Since Boltzmann factor is always positive the partition function of physical system is positive too and can not be equal to zero. The situation is changed if we alow the parameters in the hamiltonian of a system to be complex. In this case the partition function may have zeros which are called Lee-Yang zeros. Lee and Yang studied zeros of partition function for ferromagnetic Ising model with complex magnetic field \cite{Lee52}
and proved the theorem that all zeros are purely imaginary.
Latter it was proven that the Lee-Yang theorem
holds for any Ising-like model with ferromagnetic interaction \cite{Lieb81} (see
also \cite{Koz97,Koz03}).
Zeros of partition function also exist in the case when other parameters of a system are complex. In 1965 Fisher
generalized the Lee-Yang result to the case of complex temperature \cite{Fish65}.

After works of Lee, Yang  \cite{Yang52,Lee52} and Fisher  \cite{Fish65} analysis of partition function zeros are considered as a standard tool of studying properties of phase transitions in different systems \cite{Wu}. Note also that zeros of partition function fully determine the analytic properties of free energy and are very useful for studies thermodynamical properties of many-body systems. Therefore studies of partition function zeros are important fundamentally.

For a long time Lee-Yang zeros were only theoretically studied because of difficulties with realization of many-body system with complex parameters at  experiment.
Recently, in \cite{Wei12} the authors showed that experimental observation of Lee-Yang zeros of partition function of spin system is possible (see also \cite{Wei14}). The first experimental observation of Lee-Yang zeros was reported in \cite{Peng15}.

It is worth noting that
while zeros of partition function for spin systems were widely studied (see, for instance, \cite{Wei12,Peng15,Kra15,Kra16} and references therein), there are essentially smaller number of papers where zeros of partition function of Bose system (see, for instance, \cite{Mul01,Dij15,Borrmann}) and Fermi system (see, for instance,\cite{Bha11,Zvyagin}) were examined.

The present paper is inspired by the papers \cite{Wei12,Peng15}.
The aim of present paper is to relate zeros of partition function of Bose system with experimentally
observable quantities, namely, with zeros of two-time correlation functions. This relation in principle allows experimental observation of zeros of partition function of Bose system.

The paper is organized as follows. In Section 2 we find relation of zeros of two-time correlation functions for interacting Bose gas with Lee-Yang zeros. Section 3 is devoted to analysis of zeros of two-time correlation function and Lee-Yang zeros in the particular case of Bose particles on two levels.  Conclusions are presented in Section 4.

\section{Relation of zeros of two-time correlation functions with Lee-Yang zeros}

Let us consider a system of
$N$ Bose particles which is described by the following hamiltonian
\begin{eqnarray}\label{H}
H=\sum_{i=1}^s\epsilon_i \hat n_i +\gamma\sum_{i=1}^s \hat n_i^2.
\end{eqnarray}
Here  $\epsilon_i$ are energy levels of noninteracting particles, $s$ is the number of the levels, $\gamma$ is a constant of interaction ($\gamma>0$ corresponds to repulsive interaction and
$\gamma<0$ corresponds to attractive interaction),
$\hat n_i=a^+_ia_i$ is occupation numbers operator of $i$-th level with eigenvalues $n_i=0,1,2,...$, $a^+_i$, $a_i$ are creation and annihilation operators of boson on the $i$-th level which satisfy the following commutation relations
\begin{eqnarray}
[a_i,a^+_j]=\delta_{ij}.
\end{eqnarray}
 It is worth to stress that Hamiltonian (\ref{H}) can be considered as a simple variant of Bose-Hubbard
Hamiltonian \cite{Weichman,Greiner,Will}.
We consider canonical ensemble with  fixed number of particles  $N$, therefore occupation numbers $n_i$ satisfy the following condition
\begin{eqnarray}\label{cond}
\sum_{i=1}^s n_i=N.
\end{eqnarray}

Let us consider two-time correlation function of Bose system described by Hamiltonian (\ref{H})
\begin{eqnarray}\label{aa}
\langle a^+_j(t_1) a_j(t_2) \rangle=
{1\over Z}{\rm Sp} e^{-\beta H} a^+_j(t_1) a_j(t_2),
\end{eqnarray}
where $Z$ is partition function
\begin{eqnarray}
Z={\rm Sp} e^{-\beta H},
\end{eqnarray}
$\beta=1/kT$ is inverse temperature and
\begin{eqnarray}
a_j(t)=e^{iHt/\hbar}a_je^{-iHt/\hbar}.\label{at23}
\end{eqnarray}
Substituting (\ref{H}) into (\ref{at23}), we obtain
\begin{eqnarray}
a_j(t)=e^{i(\epsilon_j \hat n_j +\gamma (\hat n_j)^2)t/\hbar}
a_je^{-i(\epsilon_j \hat n_j +\gamma (\hat n_j)^2)t/\hbar}.
\end{eqnarray}
Let us rewrite $a_j(t)$ in the form which is convenient for calculation of the correlation functions. Using identity
\begin{eqnarray}
a_j f(\hat n)=a_j f(a^+_j a_j)=f(a_j a^+_j)a_j=f(\hat n+1)a_j,
\end{eqnarray}
we can write
\begin{eqnarray}
a_j(t)=e^{i(\epsilon_j \hat n_j +\gamma (\hat n_j)^2)t/\hbar} \nonumber
e^{-i(\epsilon_j (\hat n_j+1) +\gamma (\hat n_j+1)^2)t/\hbar}a_j=\\
=e^{-i(\epsilon_j +\gamma +2\gamma\hat n_j)t/\hbar} a_j.\label{at1}
\end{eqnarray}
Note that $a_j(t)$ \label{at1} can be rewritten in the following form
\begin{eqnarray}\label{at}
a_j(t)=a_j e^{-i(\epsilon_j -\gamma +2\gamma\hat n_j)t/\hbar}.\label{at2}
\end{eqnarray}
The conjugated operator to $a_j(t)$ reads
\begin{eqnarray}\label{at+}
a^+_j(t)=e^{i(\epsilon_j -\gamma +2\gamma\hat n_j)t/\hbar}a^+_j.
\end{eqnarray}
Substituting  (\ref{at}) and (\ref{at+}) into (\ref{aa}), we find
\begin{eqnarray}
\langle a^+_j(t_1) a_j(t_2) \rangle=e^{i(\epsilon_j -\gamma)\tau/\hbar}
{1\over Z}{\rm Sp} e^{-\beta H} e^{i2\gamma\hat n_j\tau/\hbar}a^+_j a_j=\nonumber\\
e^{i(\epsilon_j -\gamma)\tau/\hbar}
{1\over Z}{\rm Sp} e^{-\beta \tilde H}\hat n_j,
\end{eqnarray}
here $\tau=t_1-t_2$. We introduce effective hamiltonian
\begin{eqnarray}\label{eff}
\tilde H=\sum_{k=1}^s\tilde\epsilon_k \hat n_k +\gamma\sum_{k=1}^s \hat n_k^2
\end{eqnarray}
with
\begin{eqnarray}
\tilde\epsilon_k=\epsilon_k,\ \ {\rm when} \ \ k\ne j,\\
\tilde\epsilon_j=\epsilon_j-i{2\gamma\tau\over\beta\hbar}. \label{comp}
\end{eqnarray}
Here index $j$ is fixed and corresponds to the energy level for which we consider
the two-time correlation function (\ref{aa}).
It is important to note that effective hamiltonian given by (\ref{eff}) contains
complex parameter (\ref{comp}).

Taking $\rm Sp$ over eigenstates of occupation number operators, the two-time correlation function reads
\begin{eqnarray}\label{cor}
\langle a^+_j(t_1) a_j(t_2) \rangle=
{e^{i(\epsilon_j -\gamma)\tau/\hbar}\over Z}\sum_{n_1}\sum_{n_2}...\sum_{n_s} n_j
\exp{\left(-\beta \sum_{k=1}^s(\tilde\epsilon_k  n_k +\gamma  n_k^2)\right)}.
\end{eqnarray}
 Note, that occupation numbers in (\ref{cor}) satisfy condition (\ref{cond}). Therefore, the sum over occupation numbers in  (\ref{cor}) can not be factorized.
 The correlation function can be written in the following form
\begin{eqnarray}\label{aazero}
\langle a^+_j(t_1) a_j(t_2) \rangle=-i{\hbar\over2\gamma}{e^{i(\epsilon_j -\gamma)\tau/\hbar}\over Z}{\partial\tilde Z\over\partial\tau},
\end{eqnarray}
where the partition function $\tilde Z$ reads
\begin{eqnarray}\label{tildeZ}
\tilde Z={\rm Sp} e^{-\beta\tilde H}=\sum_{n_1}\sum_{n_2}...\sum_{n_s}
\exp{\left(-\beta \sum_{k=i}^s(\tilde\epsilon_k  n_k +\gamma  n_k^2)\right)}.
\end{eqnarray}
Here $n_k$ satisfy (\ref{cond}).
 The hamiltonian in partition function (\ref{tildeZ}) contains complex parameter and $\tilde Z$ can possess Lee-Yang zeros. It is worth mentioning that these zeros  are related with the zeros of correlation function according to (\ref{aazero}).
From (\ref{aazero}) we find
\begin{eqnarray}
\tilde Z=Z\left(i{2\gamma\over\hbar }\int_0^{\tau} d \tau'\langle a^+_j(t_2+\tau') a_j(t_2) \rangle
e^{-i(\epsilon_j -\gamma)\tau'/\hbar}  + 1\right),
\end{eqnarray}
where  we take into account that $\tilde Z=Z$ at $\tau=0$.

\section{Bose particles on two levels}

 Let us study a particular case of two-level system of $N$ Bose particles which is described by hamiltonian (\ref{H}) with $s=2$. In this case, taking into account condition (\ref{cond}), we have $n_2=N-n_1$. So, in this case correlation function (\ref{cor}) can be reduced to the following expression
\begin{eqnarray}\label{aa2}
\langle a^+_1(t_1) a_1(t_2) \rangle=
{e^{i(\epsilon_1 -\gamma)\tau/\hbar}\over Z}
e^{-\beta\epsilon_2N-\beta\gamma N^2}
\sum_{n_1=0}^N n_1
e^{-\beta( n_1(\epsilon_1-\epsilon_2-2\gamma N)+2\gamma  n_1^2)}
e^{i2\gamma n_1\tau /\hbar}.\nonumber\\
\end{eqnarray}
Here we consider correlation function for the level $j=1$ of the system.
According to (\ref{tildeZ}) partition function with complex parameter reads
\begin{eqnarray}\label{Z2}
\tilde Z=e^{-\beta\epsilon_2N-\beta\gamma N^2}
\sum_{n_1=0}^N
e^{-\beta( n_1(\epsilon_1-\epsilon_2-2\gamma N)+2\gamma  n_1^2)}
e^{i2\gamma n_1 \tau/\hbar}.
\end{eqnarray}
Partition function $Z$ has the similar expression as $\tilde Z$ but without exponent that contains complex unit.

It is convenient to introduce the complex variable
\begin{eqnarray}
q=e^{-\beta(\epsilon_1-\epsilon_2)+i2\gamma \tau/\hbar}=\rho e^{i\phi},
\end{eqnarray}
where
\begin{eqnarray}\label{rhophi}
\rho=e^{-\beta(\epsilon_1-\epsilon_2)}, \ \ \phi=2\gamma \tau/\hbar.
\end{eqnarray}
Note that $\rho$ and $\phi$ can be considered as independent variables if parameters of hamiltonian $\epsilon_1-\epsilon_2$ and $\tau$  are independent.
Then partition function $\tilde Z$ as function of $q$ reads
\begin{eqnarray}\label{Zq}
\tilde Z=e^{-\beta\epsilon_2N-\beta\gamma N^2}
P_1(q,\beta).
\end{eqnarray}
where
\begin{eqnarray}
P_1(q,\beta)=\sum_{n_1=0}^N q^{n_1}
e^{-\beta2\gamma  n_1(n_1-N)}.
\end{eqnarray}
The correlation function can be written in the form
\begin{eqnarray}\label{aa2q}
\langle a^+_1(t_1) a_1(t_2) \rangle=
{e^{i(\epsilon_1 -\gamma)\tau/\hbar}\over Z}
e^{-\beta\epsilon_2N-\beta\gamma N^2}P_2(q,\beta),
\end{eqnarray}
where
\begin{eqnarray}\label{S2S1}
P_2(q,\beta)=\sum_{n_1=0}^N n_1
q^{n_1}
e^{-\beta2\gamma  n_1(n_1-N)}=q{\partial P_1(q,\beta)\over\partial q}.
\end{eqnarray}
Here it is more convenient to use derivative over $q$ instead derivative over $\tau$ as in (\ref{aazero}).
Note that zeros of partition function and correlation function are related with polynomials $P_1(q,\beta)$ and $P_2(q,\beta)$, respectively.

Let us consider high temperature limit $\beta\to 0$.
In this case zeros of partition function are determined by the zeros of the following polynomial
\begin{eqnarray}\label{S1}
P_1(q,0)=\sum_{n_1=0}^N q^{ n_1}={q^{N+1}-1\over q-1},
\end{eqnarray}
and zeros of correlation function are determined by the zeros of
\begin{eqnarray}\label{S2}
P_2(q,0)=\sum_{n_1=0}^N n_1q^{n_1}={q\over(q-1)^2}\left(Nq^{N+1}-(N+1)q^N+1\right).
\end{eqnarray}

\begin{figure}[h!]
\begin{center}
\includegraphics[width=0.35\textwidth]{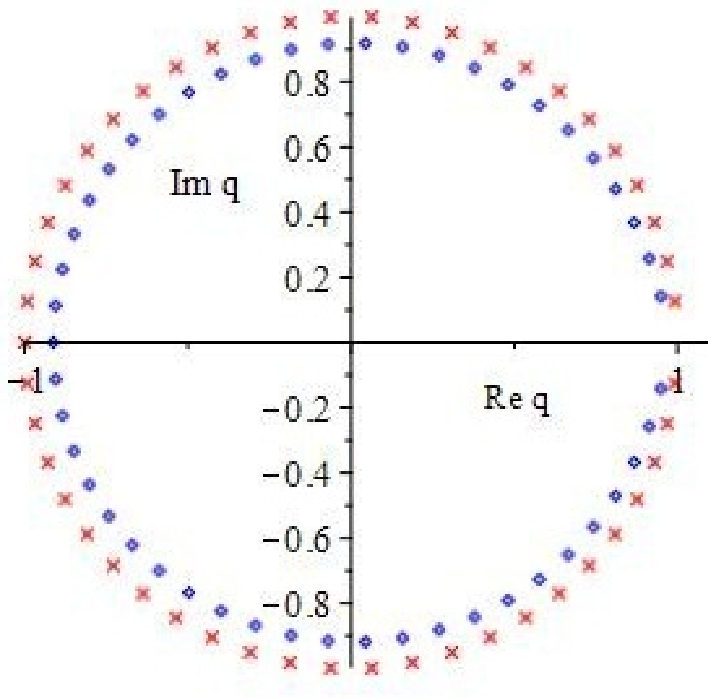}{a}
\quad \quad \quad \quad
\includegraphics[width=0.35\textwidth]{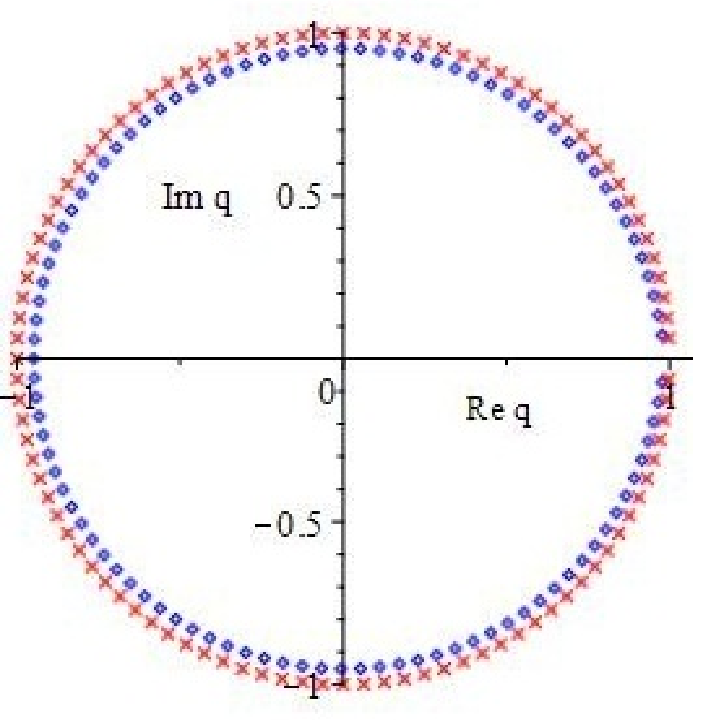}{b}
\caption{Zeros of correlation function  (marked by crosses) and Lee-Yang zeros  (marked by circles) in the limit $\beta\to 0$ for (a) $N=50$, (b) $N=100$.}
\label{f1}
\end{center}
\end{figure}

As we see from (\ref{S1}) zeros of partition function lay on the circle of unit radius
in the complex plane $q$, namely, $\rho=1$ and $\phi={2\pi l\over N+1}$, where $l=1,2,...N$.
Zeros of partition function and correlation function at $\beta\to 0$ are presented in Figure \ref{f1}.

\begin{figure}[h!]
\begin{center}
\includegraphics[width=0.35\textwidth]{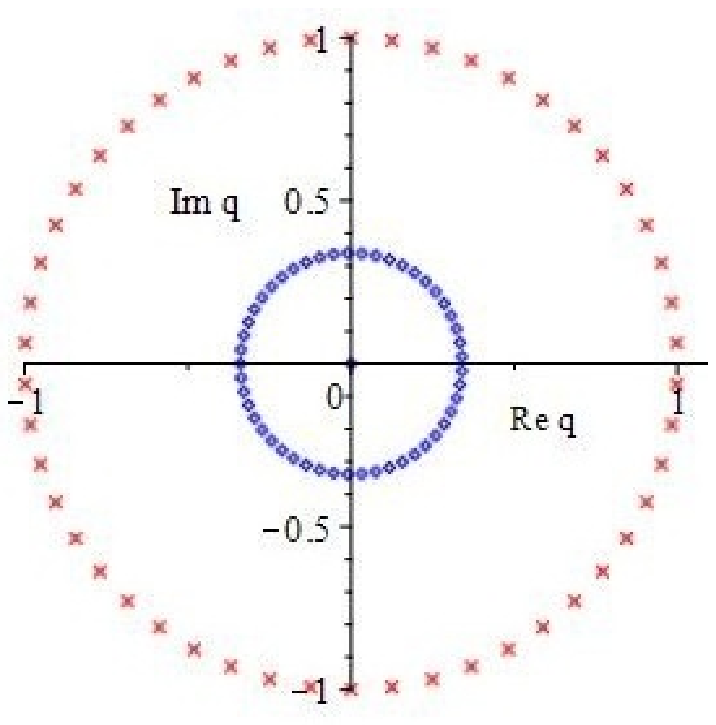}{a}
\quad \quad \quad \quad
\includegraphics[width=0.35\textwidth]{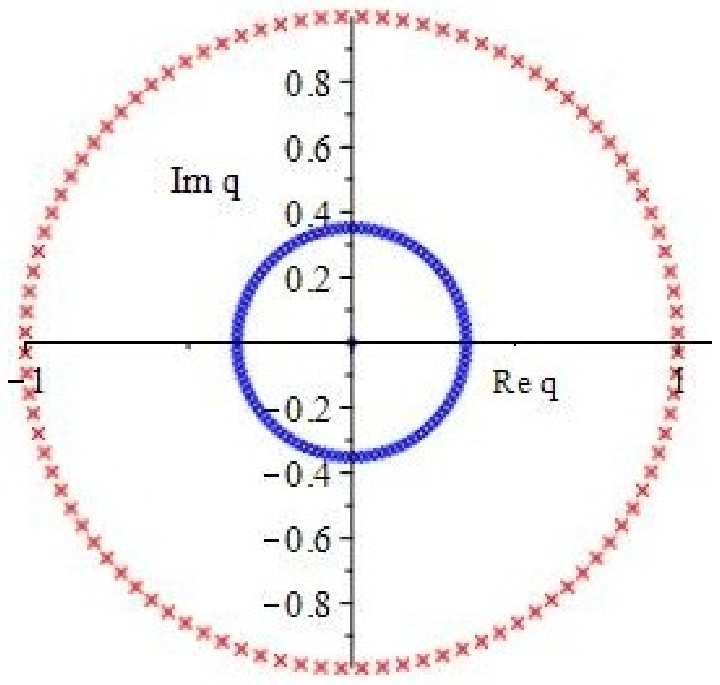}{b}
\caption{Zeros of correlation function (\ref{aa2q}) (marked by crosses) and  zeros of partition function (\ref{Zq}) (marked by circles) for  (a) $N=50$, $2\beta\gamma=-1$  (b) $N=100$, $2\beta\gamma=-1$.}
\label{f01}
\end{center}
\end{figure}

\begin{figure}[h!]
\begin{center}
\includegraphics[width=0.35\textwidth]{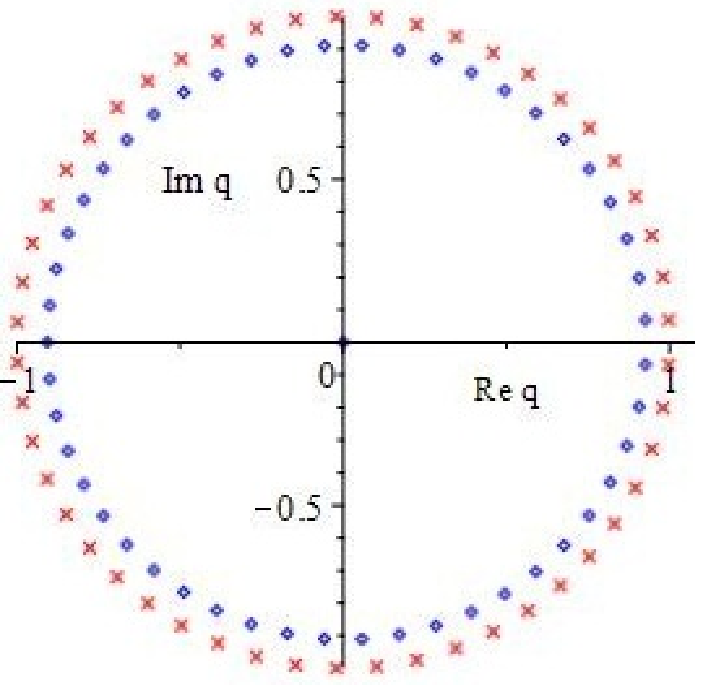}{a}
\quad \quad \quad \quad
\includegraphics[width=0.35\textwidth]{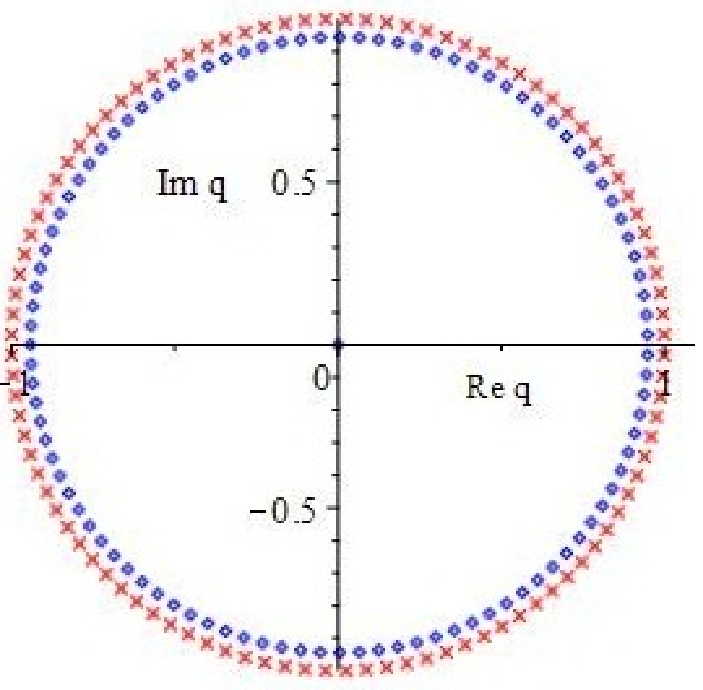}{b}
\caption{Zeros of correlation function (\ref{aa2q}) (marked by crosses) and zeros of partition function (\ref{Zq}) (marked by circles) for  (a) $N=50$, $2\beta\gamma=-0.01$  (b) $N=100$, $2\beta\gamma=-0.01$.}
\label{f02}
\end{center}
\end{figure}

\begin{figure}[h!]
\includegraphics[width=0.5\textwidth]{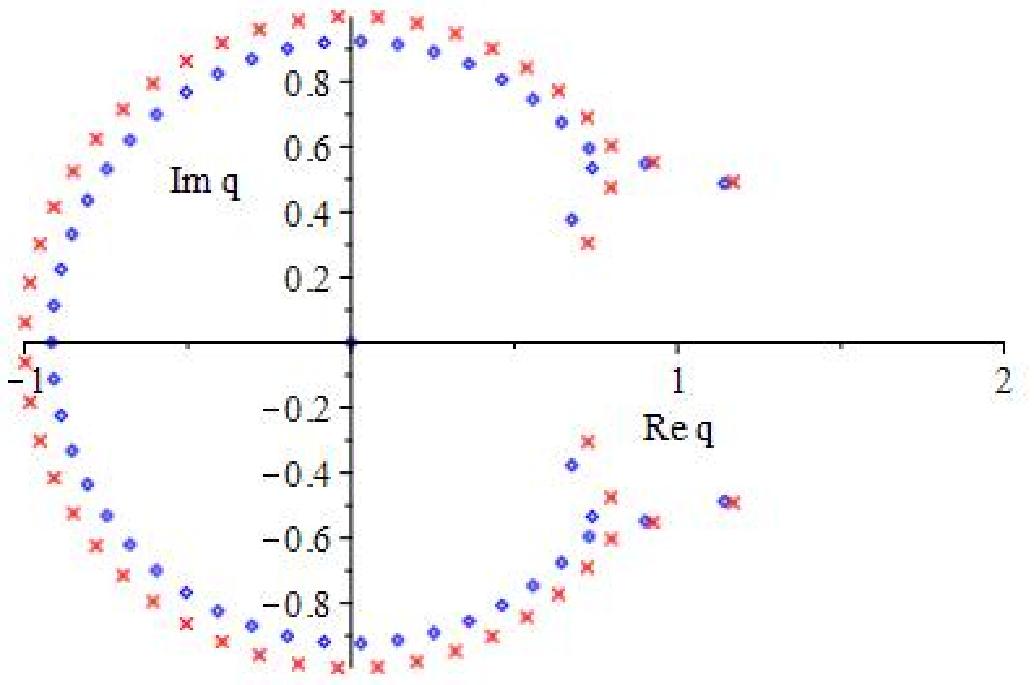}{a}
\quad
\includegraphics[width=0.5\textwidth]{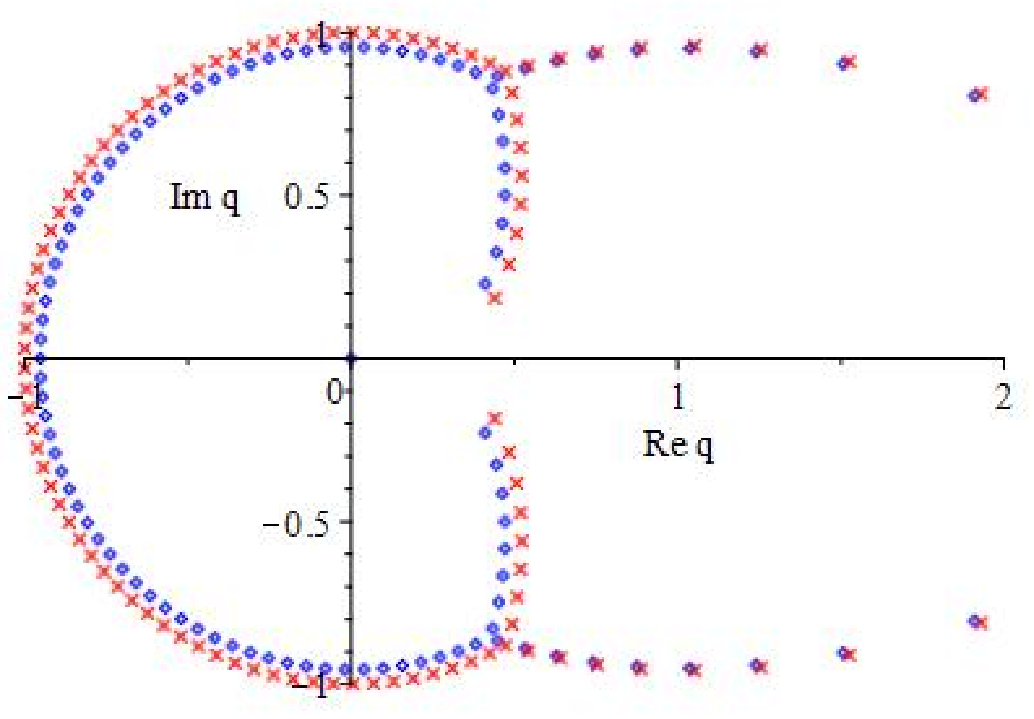}{b}
\caption{Zeros of correlation function (\ref{aa2q}) (marked by crosses) and zeros of partition function (\ref{Zq}) (marked by circles) for (a) $N=50$, $2\beta\gamma=0.01$ (b) $N=100$, $2\beta\gamma=0.01$.}
\label{f2}
\end{figure}

\begin{figure}[h!]
\includegraphics[width=0.5\textwidth]{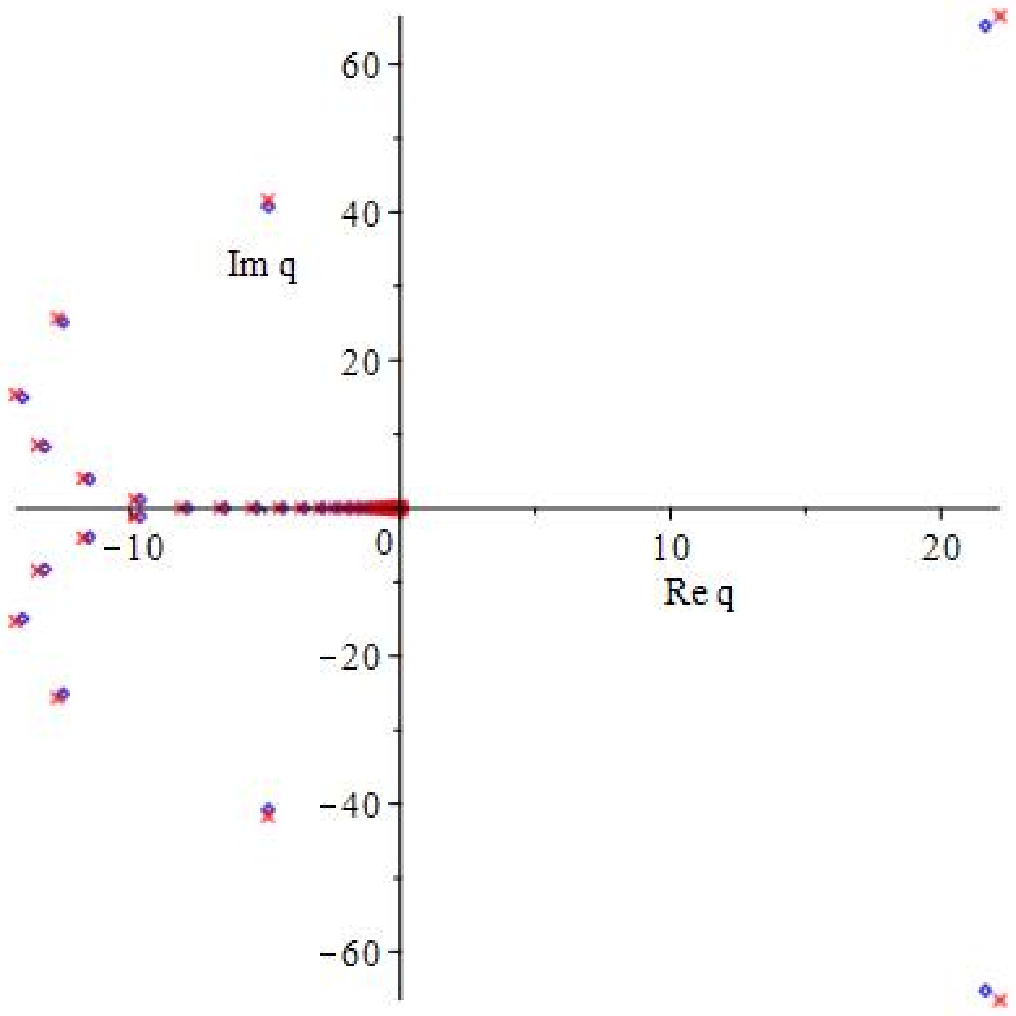}{a}
\quad
\includegraphics[width=0.5\textwidth]{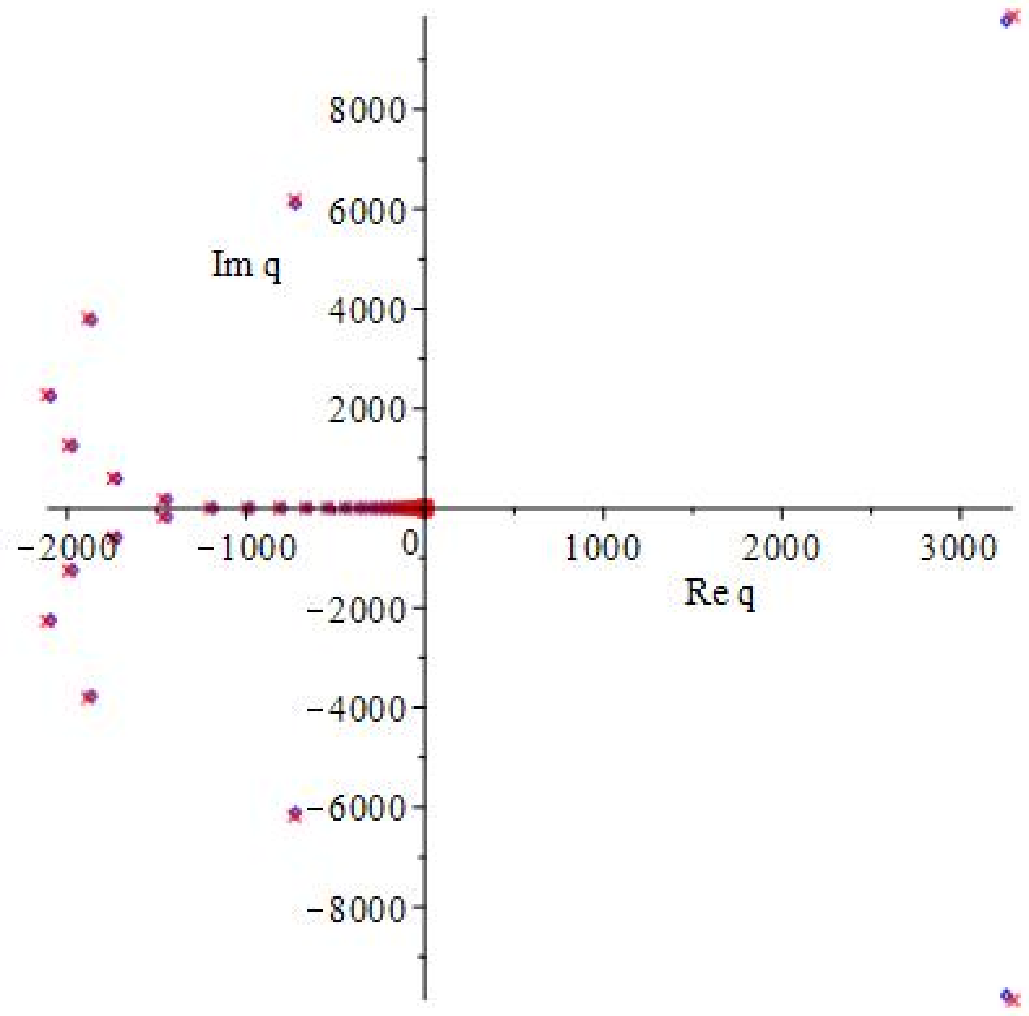}{b}
\caption{Zeros of correlation function (\ref{aa2q}) (marked by crosses) and zeros of partition function (\ref{Zq}) (marked by circles) for (a) $N=50$, $2\beta\gamma=0.1$ (b) $N=100$, $2\beta\gamma=0.1$.}
\label{f22}
\end{figure}

\begin{figure}[h!]
\includegraphics[width=0.5\textwidth]{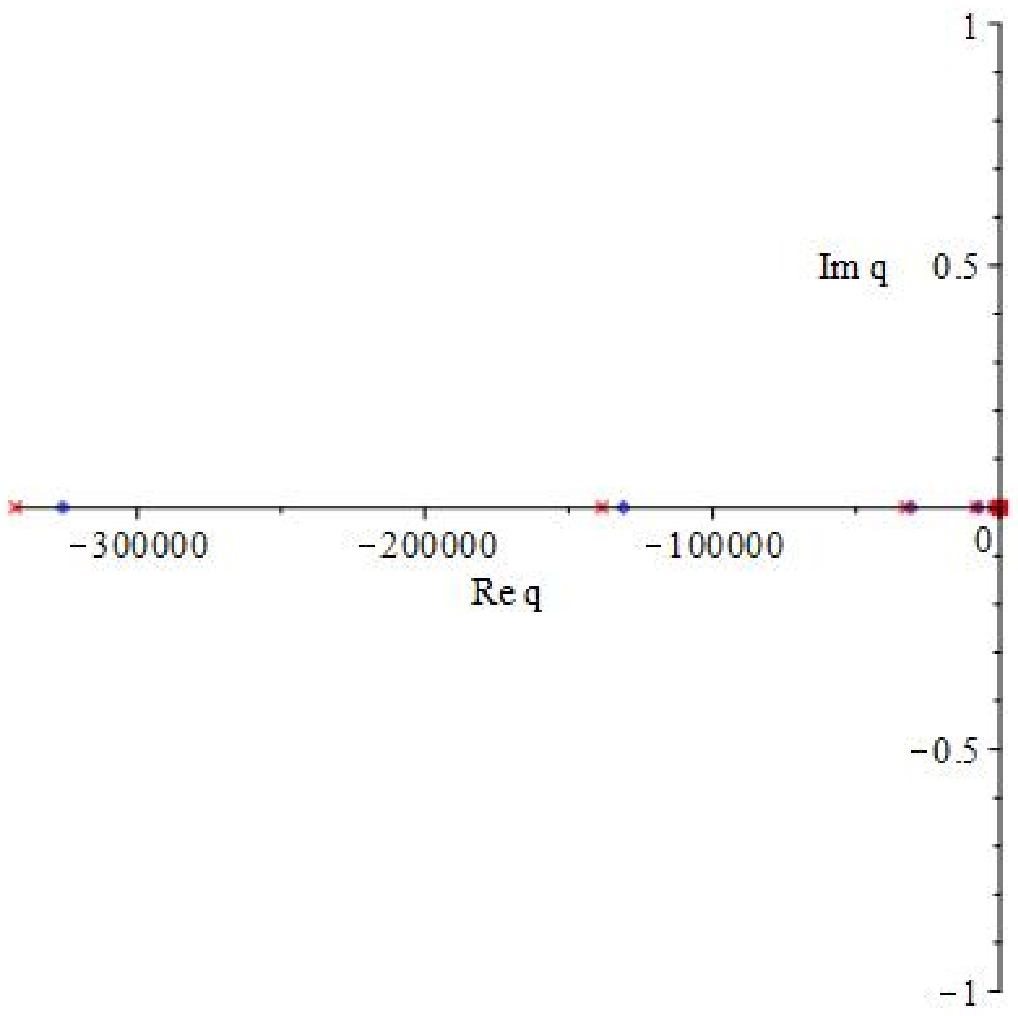}{a}
\quad
\includegraphics[width=0.5\textwidth]{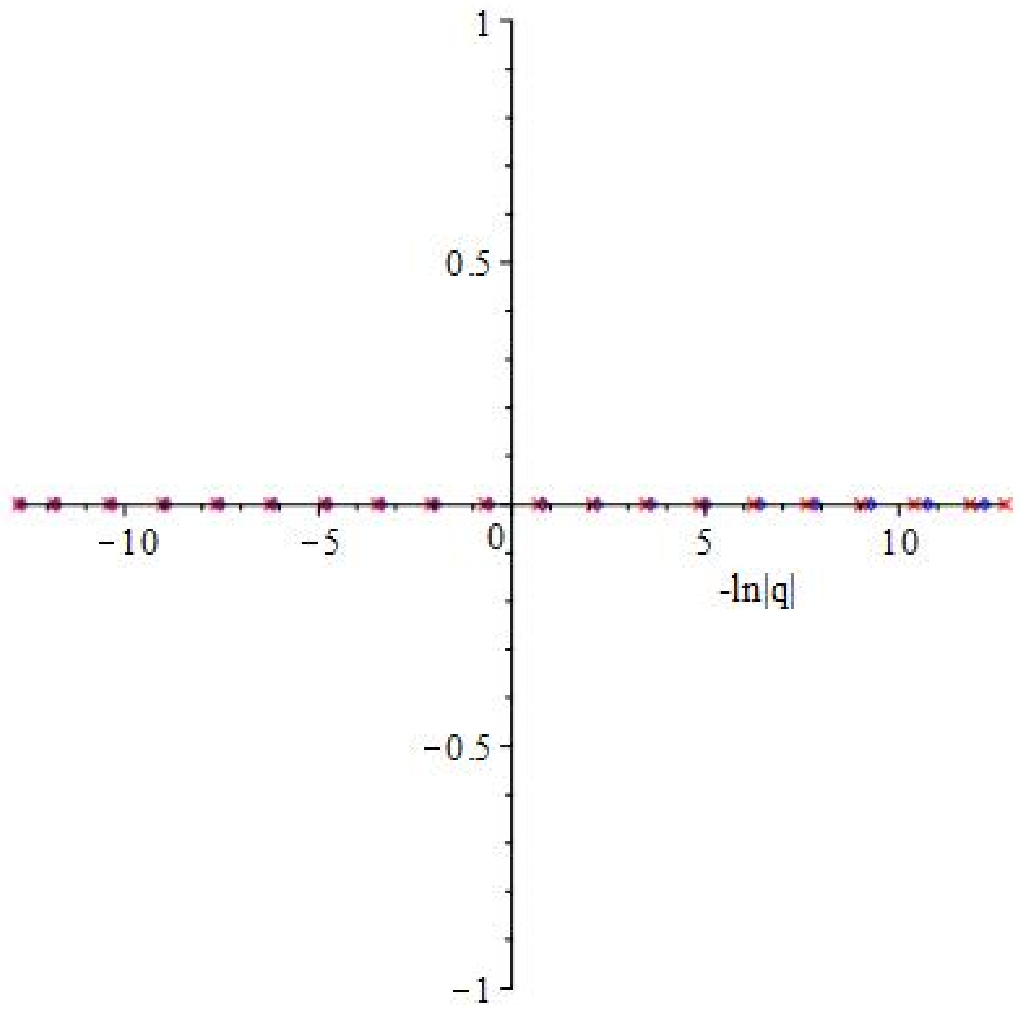}{b}
\caption{Zeros of correlation function (\ref{aa2q}) (marked by crosses) and zeros of partition function (\ref{Zq}) (marked by circles) for  $N=20$, $2\beta\gamma=\ln 2$ (a) at the complex plane q; (b) in the logarithmic scale.}
\label{f3}
\end{figure}

It is worth mentioning that  because of relation  (\ref{S2S1}) according to the Gauss-Lucas theorem the convex hull of Lee-Yang  zeros (roots
of polynomial $P_1(q,\beta)$) contains the zeros of correlation function (roots of polynomial $P_2(q,\beta)$).  This  is well seen in the Figures (\ref{f1})-(\ref{f3}), where the zeros of correlation function and Lee-Yang zeros are presented for different numbers of particles $N$ and different temperatures $\beta$.

It is also worth noting that according to the  Kurtz  theorem \cite{Kurtz} zeros of polynomial $P_n(x)=a_n x^n+a_{n-1}x^{n-1}+...+a_0$ of degree $n\geq2$ with positive coefficients are real and distinct if
\begin{eqnarray}\label{condw}
a_i^2-4a_{i-1}a_{i+1}>0, \ \ i=1,2...n-1.
\end{eqnarray}
Note that in this case the zeros  are also negative because the coefficients of polynomial are positive.
 For polynomial $P_1(q,\beta)$ condition (\ref{condw}) leads to
\begin{eqnarray}\label{condf}
\beta\gamma>\frac{\ln 2}{2}.
\end{eqnarray}
Taking into account (\ref{condw}), for polynomial $P_2(q,\beta)$ we have
\begin{eqnarray}\label{condf2}
\beta\gamma>\frac{\ln 2}{2}-\frac{1}{4}\ln\left(1+\frac{1}{(N-1)^2-1}\right).
\end{eqnarray}
As we see these conditions are satisfied for repulsive interactin ($\gamma>0$).
Note, that condition (\ref{condf2}) is stronger than (\ref{condf}). Therefore, if Lee-Yang zeros are real the zeros of correlation function are real too. This statement  also follows from the Gauss-Lucas theorem. So, in the case when condition (\ref{condf})  holds  Lee-Yang zeros and zeros of correlation function are real as it is shown in the Figure \ref{f3}.

Upper bounds for the magnitudes of all polynomial's roots can be obtained from Fujiwara bound \cite{Fuj16}. Applying it to the polynomial $P_1(q,\beta)$ for $\gamma>0$ we find that the upper bound for the magnitudes of  roots (upper bound of magnitude for roots of partition function) equals $2e^{2\beta\gamma(N-1)}$. Upper bound for the magnitudes of  $P_2(q,\beta)$ roots (upper bound for roots of correlation function) with $\gamma >0$ is $2(1-1/N)e^{2\beta\gamma(N-1)}$.
These results explain the magnitude of roots for different number of particles and different temperatures which are  presented in Figures (\ref{f2})-(\ref{f3}) corresponding to repulsive interaction.

For $\gamma <0$ the upper bound for the magnitudes of  roots for the polynomial $P_1(q,\beta)$ for large $N$ reads
${\rm max}(2e^{-2\beta|\gamma|}, 2^{(1-1/N)})$ and for $P_2(q,\beta)$ reads $2(1/N)^{1/(N-1)}e^{-2\beta|\gamma|}$.
This results explain the Figures  (\ref{f01}), (\ref{f02}) corresponding to attractive interaction.

Let us consider the limit of high inverse temperature $\beta\to\infty$  (small temperature). In this case we have
\begin{eqnarray}
P_1=1+q^N,  \ \  P_2=Nq^N.
\end{eqnarray}
So, in the limit $\beta\to\infty$ the zeros of partition function lay on the circle $q=e^{i\pi(2n+1)}$ ($n=0,1,2,...N-1$) and the set of zeros of correlation function is compressed into the point $q=0$.

\section{Conclusion}

In this paper the two-time correlation functions of interacting Bose gas  described by the Bose-Hubbard model have been studied. We have found that two-time correlation functions can be represented  as (\ref{aazero}) where partition function corresponds to a system  which is described by hamiltonian with complex parameters. These complex parameters are related with parameters of interaction in the Bose gas. As a  result, we have found relation of zeros of two-time correlation functions with Lee-Yang zeros of partition function. This relation  in principle gives the possibility to observe  Lee-Yang zeros experimentally. Note, that the Bose systems considered in this paper can be realized experimentally \cite{Greiner,Will}.

Particular case of system of interacting Bose particles on two levels has been examined and zeros of two-time correlation function of the system and Lee-Yang zeros have been studied. In the case of repulsive interaction of Bose particles we have found conditions (\ref{condf}), (\ref{condf2}) on the temperature and parameter of interaction  when zeros of partition and correlation functions are real. We conclude that zeros of correlation function and Lee-Yang zeros are real in the case of low temperatures. For sufficiently large temperatures the zeros become complex. The situation is changed drastically when
interaction in the system is attractive.
In the limit of small temperatures $\beta \to\infty$ the zeros of partition function lay on the circle $q=e^{i\pi(2n+1)}$, (here $n=0,1,2,...N-1$) and  the set of zeros of correlation function is compressed into the point $q=0$.

\section*{Acknowledgments}
This work was supported in part by the European Commission under the project STREVCOMS PIRSES-2013-612669.  The authors thank Prof. Yu. Kozitsky, Prof. Yu. Holovatch, Dr. M. Krasnytska and Dr. M. Samar for useful comments and discussions.


\begin{thebibliography}{99}
\bibitem{Yang52} C. N. Yang and T. D. Lee, Statistical Theory of Equations of State and Phase Transitions. I. Theory of Condensation, Phys. Rev. {\bf87}, 404 (1952).
\bibitem{Lee52} T. D. Lee and C. N. Yang, Statistical Theory of Equations of State and Phase Transitions.
IL Lattice Gas and Ising Model, Phys. Rev. {\bf87}, 410 (1952).


\bibitem{Lieb81} E. H. Lieb, A. D. Sokal, A General Lee-Yang Theorem for One-Component and
Multicomponent Ferromagnets, Commun. Math. Phys. {\bf 80}, 153 (I981).
\bibitem{Koz97}  Yu. V. Kozitsky, Hierarchical Ferromagnetic Vector Spin Model Possessing the Lee--Yang Property. Thermodynamic Limit at the Critical Point and Above, Journal of Statistical Physics, {\bf 87}, 799 (1997).
\bibitem{Koz03} Yu. Kozitsky, Laguerre entire functions
and the Lee–Yang property, Applied Mathematics and Computation {\bf 141}, 103 (2003).
\bibitem{Fish65} M. E. Fisher, in Lectures in Theoretical Physics, edited by
W. E. Brittin (University of Colorado Press, Boulder, CO,
1965), Vol. 7c, p. 1.
\bibitem{Wu} F. Y. Wu,  Professor C. N. Yang and Statistical Mechanics Int. J. Mod. Phys. B {\bf22}, 1899 (2008).
\bibitem{Wei12} Bo-Bo Wei and Ren-Bao Liu, Lee-Yang Zeros and Critical Times in Decoherence of a Probe Spin Coupled to a Bath, Phys. Rev. Lett. {\bf 109}, 185701 (2012).
\bibitem{Wei14} Bo-Bo Wei, Shao-Wen Chen, Hoi-Chun Po, Ren-Bao Liu, Phase transitions in the complex plane of
physical parameters, Scientific Report {\bf4} , 5202 (2014).
\bibitem{Peng15}  Xinhua Peng, Hui Zhou, Bo-Bo Wei, Jiangyu Cui, Jiangfeng Du, Ren-Bao Liu, Experimental observation of Lee-Yang Zeros, Phys. Rev. Lett. {\bf 114}, 010601 (2015).

\bibitem{Kra15} M. Krasnytska, B. Berche, Yu. Holovatch, R. Kenna, Violation of Lee-Yang circle theorem for Ising phase transitions on complex networks,  EPL, {\bf111}, 60009 (2015).
\bibitem{Kra16} M. Krasnytska, B. Berche, Yu. Holovatch, R. Kenna, Partition function zeros for the Ising model on complete graphs and on annealed scale-free networks,  J. Phys. A, {\bf 49}, 135001 (2016).
\bibitem{Mul01} O. Mulken, P. Borrmann, J. Harting, and H. Stamerjohanns, Classification of phase transitions of finite Bose-Einstein condensates in power-law traps by Fisher zeros, Phys. Rev. A, {\bf 64}, 013611 (2001).
\bibitem{Dij15} W. van Dijk, C. Lobo, A. MacDonald, and R. K. Bhaduri, Fisher zeros of a unitary Bose gas, Can. J. Phys. {\bf 93} 830 (2015).


\bibitem{Borrmann}   P. Borrmann, O. Mulken, and J. Harting, Classification of phase transitions in small systems, {\bf84}, 3511 (2000).

\bibitem{Bha11} R. K. Bhaduri, A. MacDonald and W. van Dijk, Anomalous Fisher-like zeros for the canonical partition function of noninteracting fermions, EPL, {\bf 96}, 56003 (2011).

\bibitem{Zvyagin} A. A. Zvyagin, Nonequilibrium dynamics of a system with two kinds of fermions after a pulse, Phys. Rev. B {\bf95}, 075122 (2017).
\bibitem{Weichman} M. P. A. Fisher, P. B. Weichman, G. Grinstein, D. S. Fisher,  Boson localization and the superfluid-insulator transition, Phys. Rev {\bf40}, 546 (1989).
\bibitem{Greiner} M. Greiner, O. Mandel, T. Esslinger, T. W. Hansch and I. Bloch, Quantum phase transition from a superfluid to a Mott insulator in a gas of ultracold atoms, Nature, {\bf415}, 39 (2002).
\bibitem{Will} S. Will, T. Best, U. Schneider, L. Hackermuller,  Dirk-Soren Luhmann and I. Bloch, Time-resolved observation of coherent multi-body
interactions in quantum phase revivals, Nature, {\bf465}, 197 (2010).
\bibitem{Kurtz} D. C. Kurtz, A sufficient sondition for all the roots of a polynomial to be real, The american mathematical monthly, {\bf99}, 259 (1992).
\bibitem{Fuj16} M. Fujiwara, Uber die obere Schranke des absoluten Betrages der Wurzeln einer algebraischen Gleichung, Tohoku Mathematical Journal, First series, {\bf 10}, 167 (1916).


\end{thebibliography}
\end{document}